\documentclass{iaus}
\usepackage{graphicx}
\newcommand{\oversim}[2]{\protect{\mbox{\lower0.5ex\vbox{%
   \baselineskip=0pt\lineskip=0.2ex
   \ialign{$\mathsurround=0pt #1\hfil##\hfil$\crcr#2\crcr\sim\crcr}}}}} 
\newcommand{\simgreat}{\mbox{$\,\mathrel{\mathpalette\oversim>}\,$}} 
\title[Maximal stellar mass] 
{Evidence for a fundamental stellar upper mass limit from clustered
  star formation, and some implications therof}

\author[Kroupa \& Weidner]   
{Pavel Kroupa \and Carsten Weidner}

\affiliation{Sternwarte, University of Bonn, D-53121 Bonn, Germany
\break email: pavel/cweidner@astro.uni-bonn.de}

\pubyear{2005}
\volume{227}  
\pagerange{1--10}
\date{?? and in revised form ??}
\jname{Massive Star Birth: A Crossroads of Astrophysics}
\editors{R. Cesaroni, E. Churchwell, M. Felli, \& C.M. Walmsley, eds.}
\begin{document}

\maketitle

\begin{abstract}
Theoretical considerations lead to the expectation that stars should
not have masses larger than about $m_{\rm max*}=60-120\,M_\odot$,
while the observational evidence has been ambiguous. Only very
recently has a physical stellar mass limit near $150\,M_\odot$ emerged
thanks to modern high-resolution observations of local star-burst
clusters.  But this limit does not appear to depend on metallicity, in
contradiction to theory.  Important uncertainties remain though.  It
is now also emerging that star-clusters limit the masses of their
constituent stars, such that a well-defined relation between the mass
of the most massive star in a cluster and the cluster mass, $m_{\rm
max}={\cal F}(M_{\rm ecl}) \le m_{\rm max*}\approx 150\,M_\odot$,
exists. One rather startling finding is that the observational data
strongly favour clusters being built-up by consecutively forming
more-massive stars until the most massive stars terminate further
star-formation.  The relation also implies that composite populations,
which consist of many star clusters, most of which may be dissolved,
must have steeper composite IMFs than simple stellar populations such
as are found in individual clusters.  Thus, for example, $10^5$
Taurus--Auriga star-forming groups, each with 20~stars, will ever only
sample the IMF below about $1\,M_\odot$.  This IMF will therefore not
be identical to the IMF of one cluster with $2\times 10^6$ stars.  The
implication is that the star-formation history of a galaxy critically
determines its integrated galaxial IMF and thus the total number of
supernovae per star and its chemical enrichment history. Galaxy
formation and evolution models that rely on an invariant IMF would be
wrong.

\end{abstract}

\firstsection
\section{The maximum stellar mass limit}
\label{sec1}
\subsection{A brief history}

A theoretical physical limitation to stellar masses has been known
since many decades.  Eddington (1926) calculated the limit which is
required to balance radiation pressure and gravity, the {\it Eddington
limit}: $L_{\rm Edd}/L_{\odot}\,\approx\,3.5\,\times\,10^{4}\,m /
M_{\odot}$. Hydrostatic equilibrium will fail if a star of a certain
mass $m$ has a theoretical luminosity that exceeds this limit, which
is the case for $m\simgreat 60\,M_{\odot}$. It is not clear if stars
above this limit cannot exist, as massive stars are not fully
radiative but have convective cores. But more massive stars will loose
material rapidly due to strong stellar winds.  Schwarzschild \& Harm
(1959) inferred a limit of $\approx\,60\,M_{\odot}$ beyond which stars
should be destroyed due to pulsations. But later studies suggested that
these may be damped (Beech \& Mitalas 1994).  Stothers (1992) showed
that the limit increases to $m_{\rm max*}\approx 120-150\,M_\odot$ for
more recent Rogers-Iglesia opacities and for metallicities
[Fe/H]$\approx0$. For [Fe/H]$\approx -1$, $m_{\rm max*}\approx
90\,M_\odot$. A larger physical mass limit at higher metallicity comes
about because the stellar core is more compact, the pulsations driven
by the core having a smaller amplitude, and because the opacities near
the stellar boundary can change by larger factors than for more
metal-poor stars during the heating and cooling phases of the
pulsations thus damping the oscillations. Larger physical mass limits
are thus allowed to reach pulsational instability.

Related to the pulsational instability limit is the problem that
radiation pressure also opposes accretion for proto-stars that are
shining above the Eddington luminosity.  Therefore the question
remains how stars more massive than 60 $M_{\odot}$ may be formed.
Stellar formation models lead to a mass limit near $40-100\,M_\odot$
imposed by feedback on a spherical accretion envelope (Kahn 1974;
Wolfire \& Cassinelli 1986, 1987). Some observations suggested that
stars may be accreting material in discs and not in spheres
(e.g.~Chini et al.  2004). The higher density of the disc-material may
be able to overcome the radiation at the equator of the
proto-star. But it is unclear if the accretion-rate can be boosted
above the mass-loss rate from stellar winds by this mechanism.
Theoretical work on the formation of massive stars through
disk-accretion with high accretion rates thereby allowing thermal
radiation to escape pole-wards (e.g. Nakano 1989; Jijina \& Adams
1996) indeed lessen the problem and allow stars with larger masses to
form.

Another solution proposed is the merging scenario. In this case
massive stars form through the merging of intermediate-mass
proto-stars in the cores of dense stellar clusters driven by
core-contraction due to very rapid accretion of gas with low specific
angular momentum, thus again avoiding the theoretical feedback-induced
mass limit (Bonnell, Bate \& Zinnecker 1998; Stahler, Palla \& Ho
2000).  It is unclear though if the very large central densities
required for this process to act are achieved in reality.

The search for a possible maximal stellar mass can only be performed
in massive, star-burst clusters that contain sufficiently many stars
to sample the stellar initial mass function beyond $100\,M_\odot$.
Observationally, the existence of a finite physical stellar mass limit
was not evident until very recently. Indeed, observations in the
1980's of R136 in the Large Magellanic Cloud (LMC) suggested this
object to be one single star with a mass of about
$2000-3000\,M_\odot$. Weigelt \& Baier (1985) for the first time
resolved the object into eight components using digital speckle
interferometry, therewith proving that R136 is a massive star cluster
rather than one single super-massive star.  The evidence for any
physical upper mass limit became very uncertain, and Elmegreen (1997)
stated that ``observational data on an upper mass cutoff are scarce,
and it is not included in our models [of the IMF from random sampling
in a turbulent fractal cloud]''. Although Massey \& Hunter (1998)
found stars in R136 with masses ranging up to $140-155\,M_\odot$,
Massey (2003) explains that the observed limitation is statistical
rather than physical.  We refer to this as the {\it Massey assertion},
i.e. that $m_{\rm max*}=\infty$. Meanwhile, Selman et al. (1999)
found, from their observations, a probable upper mass limit in the LMC
near about $130\,M_\odot$, but they did not evaluate the statistical
significance of this suggestion. Figer (2002) discussed the apparent
cut-off of the stellar mass-spectrum near $150\,M_\odot$ in the Arches
cluster near the Galactic centre, but again did not attach a
statistical analysis of the significance of this observation.
Elmegreen (2000) also noted that random sampling from an unlimited IMF
for all star-forming regions in the Milky Way (MW) would lead to the
prediction of stars with masses $\simgreat 1000\,M_\odot$, unless
there is a rapid turn-down {\it in the IMF} beyond several
hundred~$M_\odot$. However, he also stated that no upper mass limit to
star formation has ever been observed, a view also emphasised by
Larson (2003).

Thus, while theory clearly expected a physical stellar upper mass
limit, the observational evidence in support of this was very unclear.
This, however, changed quite dramatically only one year ago.

\subsection{R136}
Given the observed rather sharp drop-off of the IMF in R136 near
$150\,M_\odot$, Weidner \& Kroupa (2004, hereinafter WK04) studied the
{\it Massey assertion} in some detail.

R136 has an age $\le$~2.5~Myr (Massey \& Hunter 1998) which is young
enough such that stellar evolution will not have removed stars through
supernova explosions. It has a metallicity of [Fe/H]$\approx -0.5$~dex
(de Boer et al. 1985). 

From the radial surface density profile Selman et al. (1999) estimated
there to be 1350~stars with masses between 10~and $40\,M_\odot$ within
20~pc of the 30~Doradus region, within the centre of which lies R136.
Massey \& Hunter (1998) and Selman et al. (1999) found that the IMF
can be well-approximated by a Salpeter power-law with exponent
$\alpha=2.35$ for stars in the mass range 3~to $120~M_\odot$. This
corresponds to 8000 stars with a total mass of $0.68\times
10^5\,M_\odot$.  Extrapolating down to $0.1\,M_\odot$ the cluster
would contain $8\times 10^5$~stars with a total mass of
$2.8\times10^5\,M_\odot$. Using a standard IMF with a slope of
$\alpha=1.3$ (instead of the Salpeter value of 2.35) between 0.1 and
$0.5\,M_\odot$ this would change to $3.4\times 10^5$~stars with a
combined mass of $2\times 10^5\,M_\odot$, for an average mass of $0.61\,M_\odot$
over the mass range $0.1-120\,M_\odot$.
Based on the observations by
Selman et al. (1999) we assumed for our analysis that R136 has a mass
in the range $5 \times 10^{4} \le M_{\rm R136}/M_{\odot} \le 2.5
\times 10^{5}$. This mass range can be used to investigate the
expected number of stars above mass $m$,
\begin{equation}
N(>m) = \int_{m}^{m_{\rm max*}} \xi(m')\,dm',
\label{eq:Nm}
\end{equation}
with the mass in stars of the whole (originally embedded) cluster
being
\begin{equation}
M_{\rm ecl} = \int_{m_{\rm low}}^{m_{\rm max*}} m'\,\xi(m')\,dm',
\label{eq:Mecl1}
\end{equation}
where $m_{\rm low}\,=\,0.01\,M_{\odot}$ and $m_{\rm max*}=\infty$ (the
Massey assertion). Here the assumption is that the cluster is still
compact despite having-blown out its residual gas. There are two
unknowns ($N(>m)$ and $k$) that can be solved for using the two
equations above.

We used the {\it standard stellar IMF}: The distribution of stars in
clusters is well described by a multi power-law function (Kroupa
2001), $\xi(m)\propto m^{-\alpha_i}$, where $\xi(m)\,dm$ is the number
of stars in the mass interval $m,\,m+dm$.  For massive stars several
observations found the Salpeter value ($\alpha_{3} = 2.35$) for a large
variety of conditions (Massey \& Hunter 1998, Sirianni \etal\ 2000,
2002, Parker \etal\ 2001, Massey 2002, 2003, Wyse \etal\ 2002, Bell
\etal\ 2003, Piskunov \etal\ 2004). Below $0.5\,M_\odot$ the IMF
flattens (Kroupa, Tout \& Gilmore 1993, Kroupa 2001, Reid, Gizis \&
Hawley 2002), and a convenient description is
\begin{equation}
          \begin{array}{l@{\quad\quad,\quad}l}
\alpha_0 = +0.30&0.01 \le m/{M}_\odot < 0.08,\\
\alpha_1 = +1.30&0.08 \le m/{M}_\odot < 0.50,\\
\alpha_2 = +2.35&0.50 \le m/{M}_\odot < 1.00,\\
\alpha_3 = +2.35&1.00 \le m/{M}_\odot \le m_{\rm max*}.\\
          \end{array}
\label{Kroupa-IMF}
\end{equation}
In particular, Massey \& Hunter (1998) and Selman et al. (1999) found
the Salpeter power-law IMF to be valid for the R136 cluster, except
near the core where mass segregation has skewed the stellar mass
distribution towards massive stars.

\begin{figure}
\begin{center}
\includegraphics[width=8cm]{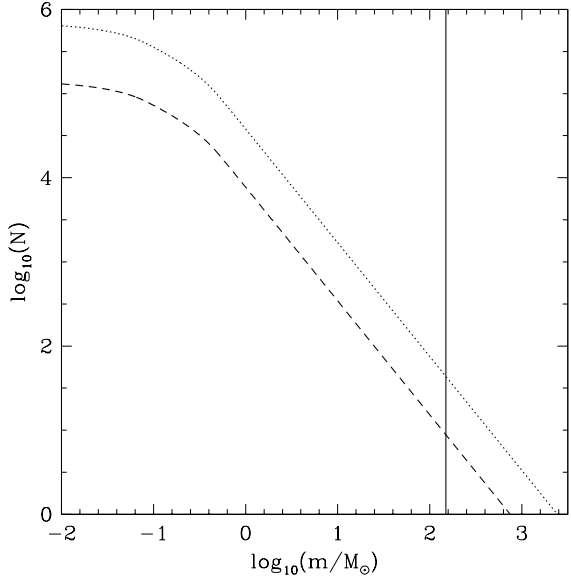}
\vspace*{-2.0cm}
\caption{Number of stars above mass $m$ for R136 with different mass
  estimates (dotted line: $M_{\rm R136} = 2.5 \times 10^5\, M_\odot$,
  dashed line: $M_{\rm R136} = 5 \times 10^4\, M_\odot$, Selman et
  al. 1999). The vertical solid line marks $m=150\,M_\odot$. Taken
  from WK04.}
\label{fig:pk_numbs}
\end{center}
\end{figure}

$N(>m)$ is plotted in Fig.~\ref{fig:pk_numbs} for the above standard IMF
and for the two mass estimates of the cluster. The solid vertical line
indicates $150\,M_{\odot}$, the approximate maximum mass observed in
R136 (Massey \& Hunter 1998).  We find that
$N(>150\,M_\odot)=40$~stars are missing if $M_{\rm ecl}=2.5\times
10^5\,M_\odot$, while $N(>150\,M_\odot)= 10$~stars are missing if
$M_{\rm ecl}=5\times 10^4\,M_\odot$. The probability that no stars are
observed although 10 are expected, assuming $m_{\rm max*}=\infty$, is
$P=4.5\times 10^{-5}$. In WK04 we concluded that the observations of the
massive stellar content of R136 suggest a physical stellar mass limit
near $m_{\rm max*}=150\,M_\odot$.

Furthermore, in WK04 we deduced that the {\it Massey assertion} would
be correct for both cluster masses if the IMF had a slope $\alpha_{3}
\simgreat 2.8$. Such a steep slope would make the observed limit
consistent with random selection from the IMF, and it may be the true
power-law index if unresolved multiple systems among O~stars are
corrected for, but this awaits a detailed study.  A further caveat
comes from unresolved multiple systems which would allow an $m_{\rm
max*, true}$ as small as $\approx m_{\rm max*}/2$ if $\alpha_3\approx
2.35$.

\subsection{Arches}
The Arches is a star-burst cluster within 30~pc in projected
distance from the Galactic centre. It has a mass $M\approx 1\times
10^5\,M_\odot$ (Bosch et al. 2001), age $2-2.5$~Myr (Najarro et
al. 2004) and [Fe/H]$\approx 0$ (Najarro et al. 2004). It is thus a
counterpart to R136 in that the Arches is metal rich and was born in a
very different environment to R136.

\begin{figure}
\begin{center}
\includegraphics[width=14cm]{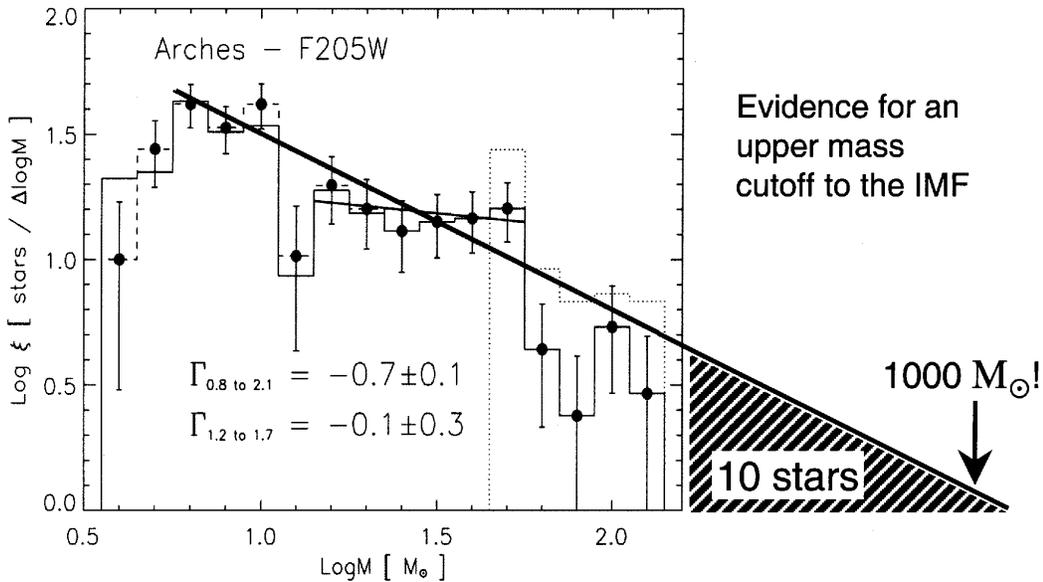}
\vspace*{-0.3cm}
\caption{Stellar mass function in the Arches cluster. Taken from Figer
  (2002) with kind permission of the author.  $\Gamma_{0.8\,{\rm
      to}\,2.1}=-(\alpha_3 - 1)$ for stars in the mass range
  log$_{10}m\in (0.8, 2.1)$.}
\label{fig:pk_figer}
\end{center}
\end{figure}

Using his HST observations of the Arches (Fig.~\ref{fig:pk_figer}),
Figer (2005) performed the same analysis as WK04 did for R136. The
Arches appears to be dynamically evolved, with substantial mass loss
through the strong tidal forces (Portegies Zwart et al. 2002) and the
stellar mass function with $\alpha=1.9$ is thus flatter than the
Salpeter IMF. Using his updated IMF measurement, Figer calculated the
expected number of stars above $150\,M_\odot$ to be 33, while a
Salpeter IMF would predict there to be 18 stars. Observing no stars
but expecting to see~18 has a probability of $P=10^{-8}$, again
strongly suggesting $m_{\rm max*}\approx 150\,M_\odot$.

\subsection{OB associations \& star clusters}
Given the importance of knowing if a finite physical upper mass limit
exists and how it varies with metallicity, Oey \& Clarke (2005)
studied the massive-star content in 9~clusters and OB associations in
the MW, the LMC and the SMC.  They predicted the expected masses of
the most massive stars in these clusters for different upper mass
limits ($120, 150, 200, 1000\,{\rm and}\,10000\,M_{\odot}$). For all
populations they found that the observed number of massive stars
supports with high statistical significance the existence of a general
upper mass cutoff in the range $m_{\rm max*}\in
(120,\,200\,M_{\odot})$,

\subsection{Concluding remarks}
The general indication thus is that a physical stellar mass limit near
$150\,M_\odot$ seems to exist. While biases due to unresolved
multiples that may steepen the IMF and/or reduce the true maximal mass
need to be studied further, the absence of variations of $m_{\rm
max*}$ with metallicity poses a problem.  A constant $m_{\rm max*}$
would only be apparent for a true variation as proposed by the
theoretical models, if metal-poor environments have a larger stellar
multiplicity, the effects of which would have to compensate the true
increase of $m_{\rm max*}$ with metallicity.

\section{Maximal stellar mass in clusters}
\begin{figure}
\begin{center}
\includegraphics[width=8cm]{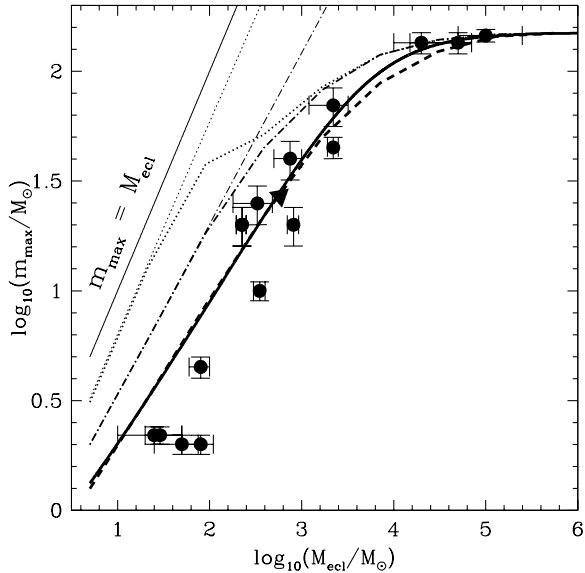}
\vspace*{-2.0cm}
\caption{The {\it thick solid line} shows the dependence of the mass
of the most-massive star in a cluster on the cluster mass according to
the semi-analytical model.  The {\it thick dashed line} shows the mean
maximum stellar mass for sorted sampling (\S~\ref{se:sosa}). The {\it
dot-dashed lines} are mass-constrained random-sampling results
(\S~\ref{se:cosa}) with a physical upper mass limit of $m_{\rm
max*}=150\,M_{\odot}$ ({\it thick line}) and $10^{6}\,M_{\odot}$ ({\it
thin line}). Pure random sampling models (\S~\ref{se:rasa}) are
plotted as {\it dotted lines}. The {\it thick} one is sampled to
$m_{\rm max*}\,=\,150\,M_{\odot}$ while the {\it thin} one up to
$10^{6}\,M_{\odot}$. The {\it thin solid line} shows the identity
relation, where a ``cluster'' consists only of one star. The {\it
dots} with error bars are observed clusters, while the {\it triangle}
is a result from a star-formation simulation with an SPH code (Bonnell
\etal\ 2003). Taken from WK05a.}
\label{fig:pk_mmaxf}
\end{center}
\end{figure}

Above we have seen that there seems to exist a universal physical
stellar mass limit. However, an elementary argument suggests that
star-clusters must also limit the masses of their constituent stars:
A pre-star-cluster gas core with a mass $M_{\rm core}$ can, obviously,
not form stars with masses $m>\epsilon\, M_{\rm core}$, where
$\epsilon\approx 0.33$ is the star-formation efficiency (Lada \& Lada
2003). Thus, given a freshly hatched cluster with stellar mass $M_{\rm
ecl}$, stars in that cluster cannot surpass masses $m_{\rm max}=M_{\rm
ecl}$, which is the identity relation corresponding to a ``cluster''
consisting of one massive star. Assuming the stellar IMF is a
continuous density distribution function and that clusters are filled
with stars distributed according to the stellar IMF, this can be
generalised by stating that each cluster can have only one most
massive star,
\begin{equation}
1 = \int_{m_{\rm max}}^{m_{\rm max*}} \xi(m')\,dm',
\label{eq:mm}
\end{equation}
with
\begin{equation}
M_{\rm ecl}(m_{\rm max}) = \int_{m_{\rm low}}^{m_{\rm max}}
m'\,\xi(m')\,dm'
\label{eq:Mecl}
\end{equation}
as a further condition, as above. These two equations need to be
solved numerically and give the semi-analytical relation $m_{\rm
max}\,=\,{\cal F}(M_{\rm ecl})$ (WK04).  It is plotted in
Fig.~\ref{fig:pk_mmaxf} as the thick-solid curve.

A literature study of clusters for which the cluster mass and the
initial mass of the heaviest star can be estimated (Weidner \& Kroupa
2005a, hereinafter WK05a) shows that the cluster mass indeed appears
to have a limiting influence on the stellar mass within it. The
observational data are plotted in Fig.~\ref{fig:pk_mmaxf}, finding
rather excellent agreement with the semi-analytical description above.

However, it would be undisputed that a stochastic sampling effect from
the IMF must be present when stars form. This can be mimicked in the
computer by performing various Monte-Carlo experiments (WK05a).  The
Monte-Carlo experiments are conducted in three different ways,
\begin{itemize}
\item[-] pure random sampling (random sampling)
\item[-] mass constrained random sampling (constrained sampling)
\item[-] mass constrained random sampling with sorted adding (sorted sampling)
\end{itemize}

\subsection{Random sampling}
\label{se:rasa}
For the random sampling 10~million clusters are randomly taken from a
cluster distribution with a power-law index of $\beta_N = 2.35$ between
12~and $2.5\times10^7$~stars. The relevant distribution function is
the embedded-cluster star-number function (ECSNF),
\begin{equation}
dN_{\rm ecl} \propto N^{-\beta_N},
\end{equation}
which is the number of clusters containing $N \in [N',N'+dN')$~stars.
Each cluster is then filled with $N$ stars randomly from the standard
IMF (eq.~\ref{Kroupa-IMF}) without a mass limit, or by imposing the
physical stellar mass limit, $m\,\le\,150\,M_{\odot}$. The stellar
masses are added to get the cluster mass, $M_{\rm ecl}$. For each
cluster the maximal stellar mass is searched for. For each cluster in
a mass bin $M_{\rm ecl} - \Delta M_{\rm ecl}/2, M_{\rm ecl} + \Delta
M_{\rm ecl}/2$ the average $m_{\rm max}$ is calculated, and the set of
average $m_{\rm max}$ values define the relation
\begin{equation}
m_{\rm max} = m_{\rm max}^{\rm ran}(M_{\rm ecl}).
\end{equation}

\subsection{Constrained sampling}
\label{se:cosa}
In this case $5\times 10^7$~clusters are randomly taken from the
embedded-cluster mass function (ECMF),
\begin{equation}
\xi_{\rm ecl}(M_{\rm ecl}) \propto M_{\rm ecl}^{-\beta}, 
\label{eq:ECMF}
\end{equation} 
 between $5\,M_{\odot}$ (the minimal, Taurus-Auriga-type, star-forming
 ``cluster'' counting $\approx$ 15 stars) and $10^{7}\,M_{\odot}$ (an
 approximate maximum mass for a single stellar population that
 consists of one metallicity and age, Weidner, Kroupa \& Larsen 2004)
 and again with $\beta=2.35$. Note that $\beta_N\approx\beta$ because
 the ECSNF and the ECMF only differ by a nearly-constant average
 stellar mass. Then stars are taken randomly from the standard IMF and
 added until they reach or surpass the respective cluster mass,
 $M_{\rm ecl}$. Afterwards the clusters are searched for their maximum
 stellar mass.  For each cluster in a mass bin $M_{\rm ecl} - \Delta
 M_{\rm ecl}/2, M_{\rm ecl} + \Delta M_{\rm ecl}/2$ the average
 $m_{\rm max}$ is calculated, and the set of average $m_{\rm max}$
 values define the relation
\begin{equation}
m_{\rm max} = m_{\rm max}^{\rm con}(M_{\rm ecl}).
\end{equation}

\subsection{Sorted sampling}
\label{se:sosa}
For the third method again $5\times10^7$~clusters are randomly sampled
from the ECMF (eq.~\ref{eq:ECMF}) between 5 $M_{\odot}$ and
$10^{7}\,M_{\odot}$ and with $\beta=2.35$. However, this time the
number $N$ of stars which are to populate the cluster is estimated
from $N=M_{\rm ecl}/m_{\rm av}$, where $m_{\rm
av}\,=\,0.36\,M_{\odot}$ is the average stellar mass for the standard
IMF (eq.~\ref{Kroupa-IMF}) between 0.01 $M_{\odot}$ and 150
$M_{\odot}$. These stars are added to give $M_{\rm ecl, ran}$,
\[M_{\rm ecl, ran} = \sum_{\rm N} m_{i},
\]
such that $m_{i} \le m_{i+1}$. If $M_{\rm ecl, ran} < M_{\rm ecl}$ in
this first step, an additional number of stars, $\Delta N$, is picked
randomly from the IMF, where $\Delta N=(M_{\rm ecl} - M_{\rm
ran})/m_{\rm av}$ (we assume $m_{\rm av}=\;$constant). Again these
stars are added to obtain an improved estimate of the desired cluster
mass,
\[^{2}M_{\rm ecl, ran} = \sum_{\rm N + \Delta N} m_{i}, \quad m_i \le m_{i+1}.
\]
This is done such that $^{2}M_{\rm ecl, ran} \approx M_{\rm ecl}$
(details of the method will be available in WK05a).  The procedure is
repeated until all clusters from the ECMF are 'filled'. They are then
also searched for the most massive star in each cluster, as above.
For each cluster in a mass bin $M_{\rm ecl}- \Delta M_{\rm ecl}/2,
M_{\rm ecl} + \Delta M_{\rm ecl}/2$ the average $m_{\rm max}$ is
calculated, and the set of average $m_{\rm max}$ values define the
relation
\begin{equation}
m_{\rm max} = m_{\rm max}^{\rm sort} (M_{\rm ecl}).
\end{equation}

\subsection{Comparison with observations}
All three relations are plotted in Fig.~\ref{fig:pk_mmaxf}. We noted
already that the observations follow the semi-analytic relation
remarkably well. Furthermore, Fig.~\ref{fig:pk_mmaxf} also suggests that
the different Monte-Carlo schemes can be selected for. Thus, the
sorted-sampling algorithm leads to virtually the same results as the
semi-analytical relation, and it fits the data very well indeed. The
correspondence of the sorted-sampling algorithm to the semi-analytical
result is not really surprising, because the algorithm is Monte-Carlo
integration of the same problem.  The constrained-sampling and
random-sampling algorithms, on the other hand, can be excluded with
very high confidence by performing statistical tests on the
observational data that are reported in detail in WK05a.
 
On a historical note, Larson (1982) had pointed out that more massive
and dense clouds correlate with the mass of the most massive stars
within them and he estimated that $m_{\rm max}=0.33\,M_{\rm
cloud}^{0.43}$ (masses are in $M_\odot$). An updated relation was
derived by Larson (2003) by comparing $m_{\rm max}$ with the stellar
mass in a few clusters, $m_{\rm max}\approx 1.2\,M_{\rm
cluster}^{0.45}$. Both are flatter than our semi-analytical relation,
and therefore do not fit the data in Fig.~\ref{fig:pk_mmaxf} as well
(WK05a). Elmegreen (1983) constructed a relation between cluster mass
and its most massive star based on an assumed equivalence between the
luminosity of the cluster population and its binding energy, for a
Miller-Scalo IMF. This function is even shallower than Larson's (2003)
relation. Assuming $m_{\rm max*}=\infty$, Elmegreen (2000) solved
eqs~\ref{eq:mm} above for a single Salpeter power-law stellar IMF
finding a $m_{\rm max}(M_{\rm ecl})$ relation quite consistent with
the data in Fig.~\ref{fig:pk_mmaxf} (WK05a).

\section{Implications}
\subsection{Stellar astrophysics and the formation of star clusters}
We are now in the happier situation that a physical stellar mass limit
seems to have been found. But the absence of clear variation of this
limit with metallicity poses a potential problem, although it may be
too early to make definite statements. Further observational work on
many more very young and massive clusters is needed to ascertain the
findings reported here, and to quantify the multiplicity properties of
massive stars, as noted above.

That our sorted-sampling algorithm for making star clusters fits the
observational maximal-stellar-mass--star-cluster-mass data so well
would appear to imply that clusters form in an organised fashion. The
physical interpretation of the algorithm (i.e. of the Monte-Carlo
integration) is that as a pre-cluster core contracts under self
gravity the gas densities increase and local density fluctuations in
the turbulent medium lead to low-mass star formation, perhaps similar
to what is seen in Taurus-Aurigae. As the contraction proceeds and
before feedback from young stars begins to disrupt the cloud,
star-formation activity increases in further density fluctuations with
larger amplitudes thereby forming more massive stars. The process
stops when the most massive stars that have just formed supply
sufficient feedback energy to disrupt the cloud. Thus, less-massive
pre-cluster cloud-cores would die at a lower maximum stellar mass than
more massive cores. But in all cases stellar masses are limited, $m\le
m_{\rm max}(M_{\rm ecl}) \le m_{\rm max*}$.

This scenario is nicely consistent with the hydrodynamic cluster
formation calculations presented by Bonnell, Bate \& Vine (2003) and
Bonnell, Vine \& Bate (2004), as is reported in more detail in WK05a.
We note here that Bonnell et al. (2004) found their theoretical
clusters to form hierarchically from smaller sub-clusters, and
together with continued competitive accretion this leads to the
relation $m_{\rm max}\propto M_{\rm ecl}^{2/3}$ in excellent agreement
with our compilation of observational data.  While this agreement is
stunning, the detailed outcome of the currently available SPH
modelling in terms of stellar multiplicities is not right (Goodwin \&
Kroupa 2005), and feedback that ultimately dominates the process of
star-formation, given the generally low star-formation efficiencies
observed in cluster-forming volumes, is not yet incorporated in the
modelling.

\subsection{Composite stellar population}
The assumption has often been made that independent of the
star-formation mode, the stellar distribution is sampled randomly from
one invariant IMF (e.g. Elmegreen 2004).  Thus, for example, $10^{5}$
clusters, each with $20$~stars, would have the same composite
(i.e. combined) IMF as one cluster with $2\times 10^{6}$~stars.

However, the existence of the $m_{\rm max}(M_{\rm ecl})$ relation has
profound consequences for {\it composite populations}.  It immediately
implies, for example, that $10^{5}$ clusters, each with $20$~stars,
{\it cannot} have the same composite (i.e. combined) IMF as one
cluster with $2\times 10^{6}$~stars, because the small clusters can
never make stars more massive than about $1\,M_\odot$.  Thus,
galaxies, that are composite stellar populations consisting of many
star clusters, most of which may be dissolved, would have steeper
composite, or integrated galaxial IMFs (IGIMFs), than the stellar IMF
in each individual cluster (Vanbeveren 1982; Kroupa \& Weidner 2003).

The IGIMF is an integral over all star-formation events in a given
star-formation ``epoch'' $t, t+\delta t$,
\begin{equation} 
\label{eq:igimf_t}
\xi_{\rm IGIMF}(m;t) = \int_{M_{\rm ecl,min}}^{M_{\rm
ecl,max}(SFR(t))} \xi\left(m\le m_{\rm max}\left(M_{\rm
ecl}\right)\right)~\xi_{\rm ecl}(M_{\rm ecl})~dM_{\rm ecl}.
\end{equation}
Thus $\xi(m\le m_{\rm max})~\xi_{\rm ecl}(M_{\rm ecl})~dM_{\rm ecl}$
is the stellar IMF contributed by $\xi_{\rm ecl}~dM_{\rm ecl}$
clusters with mass near $M_{\rm ecl}$.  $M_{\rm ecl,max}$ follows from
the maximum star-cluster-mass {\it vs}
global-star-formation-rate-of-the-galaxy relation, $M_{\rm
ecl,max}={\rm fn}(SFR)$ (eq.~1 in Weidner \& Kroupa 2005b, hereinafter
WK05b) as derived by Weidner, Kroupa \& Larsen (2004).  $M_{\rm
ecl,min}\,=\,5\,M_{\odot}$ is adopted in our standard modelling and
corresponds to the smallest star-cluster units observed.

The ``epoch'' is found by WK04 to last about $\delta t=10$~Myr; in
10~Myr we find that the embedded cluster mass function is fully
sampled, independent of the SFR. This time-scale compares very well
indeed to the star-formation time-scale in normal galactic disks
measured by Egusa, Sofue \& Nakanishi (2004) using an entirely
independent method, namely from the offset of HII regions from the
molecular clouds in spiral-wave patterns.  The time-integrated IGIMF
then follows from
\begin{equation} 
\label{eq:igimf}
\xi_{\rm IGIMF}(m) = \int_0^{\tau_{\rm G}} \xi_{\rm IGIMF}(m;t)\,dt,
\end{equation}
where $\tau_{\rm G}$ is the age of the galaxy under scrutiny.

Note that $\xi_{\rm IGIMF}(m)$ is the mass function of all stars ever
to have formed in a galaxy, and can be used to estimate the total
number of supernovae ever to have occurred, for example. $\xi_{\rm
IGIMF}(m;t)$, on the other hand, includes the time-dependence through
a dependency on $SFR(t)$ of a galaxy and allows one to compute the
time-dependent evolution of a stellar population over the life-time of
a galaxy.

Furthermore, because more-massive stellar clusters are observed to
form for higher star-formation rates SFRs (Weidner, Kroupa \& Larsen
2004), the ECMF is sampled to larger masses in galaxies that are
experiencing high SFRs, leading to IGIMFs that are flatter than for
low-mass galaxies that have had only a low-level of star-formation
activity.  WK05b show that the sensitivity of the IGIMF power-law
index for $m\simgreat 1\,M_\odot$ increases with decreasing $SFR$.
Thus, galaxies with a small mass in stars can either form with a very
low continuous SFR (appearing today as low-surface-brightness but
gas-rich galaxies) or with a brief initial SF burst (dE or dSph
galaxies), the IGIMF ought to vary significantly among such galaxies.
Low-surface-brightness galaxies would therefore appear chemically
young, while the dispersion in chemical properties ought to be larger
for dwarf galaxies than for larger galaxies (WK05b).  Another
interesting implication is that the number of supernovae per star
would be significantly smaller over cosmological times than predicted
by an invariant Salpeter IMF. 

As a general final comment, these new insights would imply that
theoretical work on galaxy formation that relies on an invariant IMF
would be wrong.

\section{Further Questions}
Unanswered questions regarding the formation and evolution of massive
stars remain. There may be stars with $m\,\ge\,m_{\rm max *}$ which
implode ``invisibly'' after 1 or 2 Myr.  The explosion mechanism
sensitively depends on the presently still rather uncertain mechanism
for shock revival after core collapse (e.g. Janka 2001).  Since such
stars would not be apparent in massive clusters older than 2~Myr they
would not affect the empirical maximal stellar mass, and $m_{\rm max*,
true}$ would be unknown at present.

Furthermore, and as stated already above, stars are often in multiple
systems. Especially massive stars seem to have a binary fraction of
80\% or even larger (Garc\'ia \& Mermilliod 2001) and apparently tend
to be in binary systems with a preferred mass-ratio near unity. Thus,
if all O~stars would be in equal-mass binaries, then $m_{\rm max
*\,true} \approx m_{\rm max *}/2$.

Finally, it is disconcerting that $m_{\rm max
  *}\,\approx\,150\,M_{\odot}$ appears to be the same for
low-metallicity environments ([$\rm Fe/H$] = -0.5, R136) and
metal-rich environments ([$\rm Fe/H$] = 0, Arches), in apparent
contradiction to the theoretical values (Stothers 1992). Clearly, this
issue needs further study.

\begin{acknowledgments}
We thank the organisers for a splendid and most stimulating meeting.
This research has been supported by DFG grant KR1635/3.
\end{acknowledgments}


\end{document}